
\magnification=\magstep1
\baselineskip=16pt
\vsize=8.9 true in
\hsize=6.5 true in
\hoffset=-0.1truecm\voffset=0.3truecm
\nopagenumbers\parindent=0pt
\footline={\ifnum\pageno<1 \hss\thinspace\hss
    \else\hss\folio\hss \fi}
\pageno=-1

\newdimen\windowhsize \windowhsize=13.1truecm
\newdimen\windowvsize \windowvsize=6.6truecm

\def\ie{{\it i.e.,~}}
\def\de{\partial}
\def\heading#1{
    \vskip0pt plus6\baselineskip\penalty-250\vskip0pt plus-6\baselineskip
    \vskip2\baselineskip\vskip 0pt plus 3pt minus 3pt
    \centerline{\bf#1}
    \global\count11=0\nobreak\vskip\baselineskip}
\count10=0
\def\section#1{
    \vskip0pt plus6\baselineskip\penalty-250\vskip0pt plus-6\baselineskip
    \vskip2\baselineskip plus 3pt minus 3pt
    \global\advance\count10 by 1
    \centerline{\expandafter{\number\count10}.\ \bf{#1}}
    \global\count11=0\nobreak\vskip\baselineskip}
\def\subsection#1{
    \vskip0pt plus3\baselineskip\penalty-200\vskip0pt plus-3\baselineskip
    \vskip1\baselineskip plus 3pt minus 3pt
    \global\advance\count11 by 1
    \centerline{{\it {\number\count10}.{\number\count11}\/})\ \it #1}}
\def\firstsubsection#1{
    \vskip0pt plus3\baselineskip\penalty-200\vskip0pt plus-3\baselineskip
    \vskip 0pt plus 3pt minus 3pt
    \global\advance\count11 by 1
    \centerline{{\it {\number\count10}.{\number\count11}\/})\ \it #1}}
\def\etal{{\it et al.\/}}
\def\eol{\hfil\break}
\def\affl#1{\noindent\llap{$^{#1}$}}
\def\simlt{\lower.5ex\hbox{$\; \buildrel < \over \sim \;$}}
\def\simgt{\lower.5ex\hbox{$\; \buildrel > \over \sim \;$}}

{
\def\cl#1{\hbox to \windowhsize{\hfill#1\hfill}}
\hbox to\hsize{\hfill\hbox{\vbox to\windowvsize{\vfill
\bf
\cl{INSTABILITIES IN PHOTOIONIZED}
\cl{INTERSTELLAR GAS}
\bigskip
\cl{Edvige~Corbelli and Andrea~Ferrara}
\bigskip\rm
\cl{Preprint n.~8/95}

\vfill}}\hfill}}

\vskip5truecm
{\leftskip1.7truecm
\affl{}Osservatorio Astrofisico di Arcetri,
\eol
Largo E.~Fermi 5, I-50125 Firenze (Italy)

\vfill
To appear in The Astrophysical Journal (July 10, 1995).
\vglue3truecm
}
\eject

\vglue 5 true cm
\heading{ABSTRACT}

\vglue 0.2 true in

We present a linear analysis of acoustic and thermo-reactive
instabilities in a diffuse gas, photoionized and heated by a radiation
field, cooled by collisional excitation of hydrogen and
metal lines.
The hydrogen recombination reaction has a stabilizing effect
on the thermal mode found by Field (1965) since the condensation
instability occurs in a narrower region of the parameter space and
grows on longer time scales due to its oscillatory character.
This effect is stronger when the mean
photon energy is not much larger than the hydrogen ionization energy.
Moreover, for fixed values of thermal pressure and photoionization
rate, there are thermo-reactive unstable equilibria
for which no transition to a stable phase is possible.
By extending our analysis of the thermo-reactive modes to the nonlinear regime
we show that when no phase transition is possible the medium
evolves through a series of nonequilibrium states characterized
by large amplitude, nonlinear periodic oscillations of temperature,
density and hydrogen ionization fraction.
We find also unstable acoustic waves which,
for solar metal abundances, are the fastest growing modes in
two temperature intervals: around  $T\sim 100$~K and $T\sim 8000$~K
(\ie cold and warm phase, respectively), independent of the mean
photon energy.
Possible implications for the interstellar medium and intergalactic medium
are briefly outlined.

\medskip
\noindent\underbar{\strut Subject}\ \underbar{\strut headings}:
Instabilities -- Interstellar matter.

\vglue\windowvsize plus 1fill minus \vsize
\eject

\pageno=1
\section{ INTRODUCTION}

\vglue 0.1 true in

Diffuse gas irradiated by an external radiation field is
commonly found in many astrophysical environments. Often the
radiation field constitutes the main energy input for the medium
through ionization processes. For example, the cosmic UV background
is responsible for heating and ionizing the Ly$\alpha$
clouds (Ikeuchi \& Turner 1991; Miralda-Escud\'e \& Ostriker 1992;
Charlton \etal 1993; Kulkarni \& Fall 1993) and the
outer disks of galaxies (Corbelli \& Salpeter 1993a,b;
Maloney 1993; Dove \& Shull 1994). Photoionization models of the
Broad Line Region have been successful in the interpretation of AGN
spectra (Kwan \& Krolik 1981; Collin-Souffrin 1990).  Finally, in our
Galaxy, examples of objects in which photoionization processes play
a primary role are the High Velocity Clouds (HVCs) (Songaila \etal
1989; Ferrara \& Field 1994), and the extended electron layer
(Reynolds 1993, Domg\"orgen \& Mathis 1994). In this paper we shall be
interested in mostly neutral gas, and our results do not
apply to objects like bright HII regions, where cooling
processes are likely be different from the ones considered here.

Interstellar gas very often tends to develop a multi-phase structure
caused by ongoing thermal instabilities (Field 1965; Field
\etal 1969; McKee \& Ostriker 1977) and evidence of a two-phase medium
has been found also in some of these photoionized diffuse objects
(see for example Ferrara \& Field 1994 and references therein, Krolik
\etal 1981). In addition, gas at high galactic latitude and in outer
disks of galaxies shows broad HI emission profiles. Broad features might
originate from a turbulent medium or from a kinetic
temperature. In both cases the phenomenon is of interest since
in outer disks there is no evidence of substantial star forming activity, which
could be responsible for the turbulence, and temperatures derived for
a uniform warm medium are often thermally unstable against temperature
and density perturbations (Dickey, Murray, \& Helou 1990; Verschuur \& Magnani
1994).
In this respect, we believe it is important to consider the
effect of ionization processes on the
thermal instability and other unstable modes in diffuse media.

A new mode,
in addition to the usual thermal (or condensation) mode and to the
wave (or acoustic) modes, is found when the time dependence of ionization is
taken into account. We shall refer to this new mode and to the
thermal one together as thermo-reactive modes.
The ionization and recombination reaction has a stabilizing
effect on the thermal instability of a pure hydrogen gas which is
collisionally ionized (Defouw 1970; Iba\~nez \& Parravano 1983;
Yoneyama 1973). The open question is whether the same effect arises in a
photoionized gas or in gas where cooling is dominated by metal line
excitation.
Goldsmith (1970) worked out a first approach to the problem
by studying the transition between the thermally unstable and stable
regimes for a metal cooled medium, photoionized and heated by an external
radiation field. Flannery \& Press (1979) pointed out the presence of
unstable acoustic modes in the thermally stable cold phase of this
gas, but in general our knowledge of the properties and existence of
acoustic and thermo-reactive instabilities is still fragmentary.

Photoionized gas in many objects is mostly warm or in a
multi-phase structure. In this paper we will examine in detail
both thermo-reactive and acoustic modes over a wider range of
temperatures, metallicity and photon fluxes. Furthermore, the analysis of
thermo-reactive instabilities will be extended to the nonlinear regime.
We neglect the magnetic field; when present, it implies a reduction of the
thermal conduction coefficient and of the growth rate of the perturbations
in the direction perpendicular to the field lines.

The plan of the paper is as follows: in \S~2 we present the basic
equations for the equilibrium and stability analysis; in \S~3 we
concentrate on the properties of acoustic
and thermo-reactive modes. Detailed numerical results of the linear analysis
are shown in \S~4, and \S~5 contains the results of the nonlinear analysis for
the thermo-reactive modes. Possible applications and a brief
summary of the most interesting results are given in \S~6.

\vglue 0.1 true in

\section
{PHYSICAL PROCESSES AND BASIC EQUATIONS}

\vglue 0.1 in

We consider an ideal homogeneous gas with metallicity $Z$, and
ratio of specific heats $\gamma$. We assume that hydrogen is the only reacting
species. The basic equations are
$${d\rho\over dt}+\rho{\bf \nabla\cdot}{\bf v}=0,\eqno(2.1)$$
$$\rho{d{\bf v}\over dt}+{\bf\nabla}p =0,\eqno(2.2)$$
$$N_0 {dx\over dt}- I(x,\rho,T)=0, \eqno(2.3)$$
$${N_0 \over \gamma -1} (1+x) k_B {dT\over dt}+N_0\Bigl({k_BT\over \gamma-1}
+\chi\Bigr){dx\over dt} + {\cal L}(x,\rho,T)-{p\over \rho^2}{d\rho\over dt}=0,
\eqno(2.4)$$
$$p=N_0\rho (1+x) k_B T,\eqno(2.5)$$

\noindent
where the various symbols are defined in Table 1. Equations (2.1) and (2.2)
express mass and momentum conservation; equation (2.3) describes the
change of the hydrogen ionization fraction, $x$. In the energy
equation (2.4)  the first two terms describe changes of the internal energy
due to variations of the temperature and of the degree of ionization,
respectively. The binding energy $\chi$ of
electrons to H atoms has been explicitly included in the internal energy,
therefore we shall also take it into account in
the heat-loss function ${\cal L}$.
In this paper we neglect magnetic fields and
heat conduction and use $\gamma=5/3$.

The steady state equilibrium is characterized by the values of
$\rho_0, x_0, T_0, p_0$ and by
$v_0 = 0$; as usual ${\cal L}(x_0, \rho_0, T_0)= I(x_0, \rho_0, T_0) = 0$.
We first investigate the response of the system to
infinitesimal perturbations of the form

$$\delta f=f_1 e^{nt+i{\bf {k\cdot r}}};\eqno(2.6)$$

\noindent
where $n=n_r+i n_i$ is the growth rate; thus $n_r > 0$ ($n_r< 0$)
means that the mode is unstable (stable).
The perturbed variables $\rho,v,x,T,p$ will be then written as
$f_0+f_1$. Linearizing the system (2.1)-(2.5), we obtain the equivalent
homogeneous system $M\zeta=0$,
where the vector $\zeta$ and the matrix $M$ are

$$\zeta\equiv\left(\matrix {\rho_1\cr v_1\cr x_1\cr T_1\cr p_1\cr}\right);$$

$$M\equiv\left(\matrix {n& ik\rho_0& 0& 0& 0\cr 0& \rho_0 n& 0& 0& ik\cr
-I_{\rho}&
0& N_0n-I_{x}& -I_{T}& 0\cr -{1\over \rho_0}& 0& -{1\over (1+x_0)}& -
{1\over T_0}& {1\over p_0}\cr
-{p_0\over \rho_0^2}n+{\cal L}_{\rho}& 0& N_0({3\over 2} k_B T_0+
\chi)n+{\cal L}_x&
{3\over 2}N_0(1+x_0)k_Bn+{\cal L}_T& 0\cr}\right)$$

\noindent
The above system has a non-trivial solution only if $\vert M\vert$=0, which
gives
the following characteristic equation for the growth rate $n$

$$\sum_{j=0}^{4}a_j n^{4-j}=0, \eqno(2.7) $$

\noindent
where
$$ a_0=1; \qquad a_1= c_s\Bigl\lbrack K_T +
(1+\tilde\chi)J_T-J_x\Bigr\rbrack;\qquad
a_2=c_s^2\Bigl\lbrack K_xJ_T-K_TJ_x+k^2\Bigr\rbrack;$$

$$a_3=k^2c_s^3\Bigl\lbrack{3\over5}(K_T-K_{\rho})+\Bigl(1+{3\over5}\tilde\chi\Bigr)
(J_T-J_\rho)-(J_x-J_\rho)\Bigr\rbrack; $$
$$a_4={3\over 5}k^2c_s^4\Bigl\lbrack(K_x-K_\rho)(J_T-J_\rho)
-(K_T-K_\rho)(J_x-J_\rho)\Bigr\rbrack.$$

In the definition of $K_T,K_\rho, K_x,J_T,J_\rho,J_x$ given in Table 1 we have
introduced some characteristic time scales of the system: three cooling
times $\tau_T$, $\tau_\rho$, $\tau_x$, and three reactive times
$\bar\tau_T$, $\bar\tau_\rho$, $\bar\tau_x$. Various combinations of
these time scales, together with the dynamical time $\tau_d$, characterize the
evolution of the perturbed gas.
The isochoric regime holds when the cooling and reactive times are much smaller
than the dynamical time,
while in the isobaric regime the dynamical time is the shortest timescale.
The general solution of a fourth degree polynomial as in eq. (2.7) is known
and can be obtained analytically; however, due to its complexity, it is not
very handy. A criterion, derived by
Hurwitz (see for example Aleksandrov \etal 1963), provides a
necessary and sufficient condition, in terms of the
$a_j$, for the real part of the roots to be all negative,
\ie stable modes for our case. This criterion has been applied
by Yoneyama (1973) to a fourth degree dispersion relation
similar to that given by equation (2.7) both for
small and large wavenumbers.

We consider a medium heated and ionized by a background radiation field
of mean photon energy $E_0$ (weighted by the photoionization cross
section) and ionization rate $\xi_0$.
$E$ and $\xi$ are the values of these quantities at a given optical
depth, and in this paper we neglect any radiative transfer effects.
We are interested in low density gas with equilibrium
temperature $30<T<30,000$~K; we neglect three-body recombinations,
and radiative recombinations are considered to levels $\ge 2$
(on-the-spot approximation) with a rate coefficient $\alpha$.
Collisional ionization can become important at high temperature
and we have adopted the ionization rate $\gamma_c$ given by
Black (1981).
Secondary electrons are included according to
the results of Shull \& Van Steenberg (1985); the number of secondary
electrons $\phi$, and the heat released for each photoionization, $E_h$,
are functions of the hydrogen ionization fraction and of the
photon energy $E$.
We examine models for different values of the metallicity $Z$,
ranging from the pure hydrogen case ($Z=0$) to metal rich systems
with $Z=2$.
The adopted ionization-recombination function $I$ and heat-loss function
${\cal L}$ per gram are the following

$$I=N_0^2 \rho \Biggl\lbrace {(1-x)\over N_0 \rho} \xi (1+\phi)
- x^2\alpha+(1-x)x\gamma_c\Biggr\rbrace \eqno(2.8)$$

$${\cal L}=N_0^2\rho\Biggl\lbrace (1-x)Z\Lambda_{HZ}+xZ\Lambda_{eZ}+
(1-x)x\Lambda_{eH}+x^2\Lambda_{eH^+}-{(1-x)\over N_0\rho} \xi
[E_h+(1+\phi)\chi] \Biggr\rbrace\eqno(2.9)$$

The energy losses due to recombinations require some care in
the calculation when we include $\chi$, the binding energy of the
electrons in the ground state n=1. The final expression for
$\Lambda_{eH^+}$ in the on-the-spot approximation is

$$\Lambda_{eH^+}=(E^{th} +\chi)\alpha;\eqno (2.10)$$

\noindent
$E^{th}$ is the thermal energy lost by recombination. Using equations
(2.10) and (2.3) it
would be possible to rewrite the energy equation (2.4)
only in terms of thermal energies if collisional ionizations are negligible,
but we prefer this form for the sake of clarity and completeness.
The functions $E^{th}$ and $\alpha$ have been adapted
from Seaton (1959)

$$\alpha={2.06\times 10^{-11}\over \sqrt{T}} (0.50\ln\Theta+
0.47\Theta^{-1/3}-0.32)~{\rm cm^3~~ s^{-1} }; $$

$$E^{th}\alpha=7.2\times 10^{-30} T \sqrt{\Theta}(0.5
\ln\Theta+0.64\Theta^{-1/3}-0.821) {\hbox{~~erg~~ cm}}^3 {\rm ~~s}^{-1}; $$

Note that even if we include recombinations to
the ground state in the above expression, the cooling is considerably
larger than the one used by Defouw (1970, cfr. eq.~[22]) for recombinations
to the n=1 level only: at $T=4000$ K, for example, the cooling due to
recombinations to all levels
is a factor 2.4 larger than that due to recombination to n=1 only.
Moreover, if the on-the-spot approximation is not used, and ionization
by an external radiation field is considered, Lyman continuum photons
should be added in the ionization equation.

It turns out that in our models cooling due to recombinations is
ineffective compared to the other types of cooling.
The expressions for energy losses due to collisional excitation of
line radiation have been taken from  Dalgarno \& McCray (1972), except
for the fine structure excitation of CII and OI by atomic hydrogen impact, for
which we used the results of Launay \& Roueff (1977), and for the Ly$\alpha$
excitation of neutral hydrogen  for which we refer to Spitzer (1978).
The metal species considered are CII, SiII, FeII, SII, OII, NI,
OI; we have assumed that the ionized to neutral ratios for O and N are the same
as for hydrogen to account for charge exchange effects.

\vglue 0.1 true in

\section
{ACOUSTIC AND THERMO-REACTIVE INSTABILITIES}

\vglue 0.1 in

In Figure 1 we present the equilibrium curves in the plane
$p_0/\xi-T_0$ for four different values of the metallicity $Z=0,
10^{-3}, 0.5, 1$ and three values of the mean photon energy $E=
15, 40, 100$~eV taken as representative of gas at different optical
depth. The dashed line segments indicate unstable equilibria;
\ie at least one of the four modes has positive $n_r$ for some $k$.
The specific nature of the various unstable modes will be discussed
later on in this Section.

For the pure hydrogen case ($Z=0$), the equilibrium curves are
monotonically decreasing functions of temperature and are generally
stable. The cooling function is dominated by Ly$\alpha$
line emission excited by e-H impact; free-free cooling is negligible
at these temperatures as well as recombination radiation losses since
for a photoionized medium these are important only at temperatures
$T\sim E_h/k_B$ (see also Section 4.3). As the mean energy of the
penetrating photons increases, the heat deposited per photoionization
gets larger and a higher density is needed to balance the heating
via collisional cooling. Note that for the case $Z=0$,
becuse of the lack of metal coolants, below 6500~K
no realistic equilibria exist.
Due to the low efficiency of the e-H impact cooling below 6500~K,
even a small amount of metals, as for $Z=10^{-3}$, makes the metal
cooling dominate at these temperatures, and a three-phase medium is
found for $E\gg 13.6$ eV. For photon
energies slightly above 13.6~eV each photoionization releases
no more than a few eV and an enhancement of the metal cooling has
two consequences: (a) the maximum possible equilibrium gas temperature
decreases; (b) $p_0/\xi$ is a monotonically decreasing function of $T_0$,
{\it \ie  multi-phase equilibria cannot exist}.

A consequence of the Field (1965) isobaric instability criterion
$(\de {\cal L}/\de T)_p< 0$, is
that  thermally unstable regions are characterized or
by a positive slope of the $p_0-T_0$ curve.
This follows from the following relation
$$\left({d\ln p\over d\ln T}\right)_{\cal L}=-{T\over \rho}{(\de {\cal L}/\de
T)_p\over
(\de {\cal L}/\de \rho)_T},$$
being  $(\de {\cal L}/\de \rho)_T>0$ for most astrophysical applications.
When ${\cal L}$ depends on $x$,
as well as on $\rho$ and $T$, there is an additional term
proportional to the temperature derivative of the ionization fraction
$$\left({\de {\cal L}\over \de T}\right)_p=\left({\de {\cal L}\over \de
T}\right)_{\rho,x}
-{\rho\over T}\left({\de {\cal L}\over \de \rho}\right)_{T,x}+{\de x\over \de
T}
\left[{\de {\cal L}\over \de x}-{\de {\cal L}\over \de \rho}{\rho\over
1+x}\right]_
{\rho,T}<0$$
or, in our notation,
$$(K_T-K_\rho)+{\de x\over \de T}(K_x-K_\rho)<0;$$
this criterion reduces to the Field criterion when $x$ is kept constant;
otherwise one cannot read  the stability properties directly
off the slope of the equilibrium curve.

The low photon energy cases in Figure 1 exemplify very well the above
point since for $Z=1$ there are negative slope parts of the equilibrium curve
that show unstable isobaric modes, contrary to what would be expected applying
Field criterion. As we raise the photon energy, the slope of $p_0/\xi$
becomes positive for a certain range of temperatures. For
photon energies above 100~eV the curves are similar to that
of $E=100$~eV, but they are shifted towards larger values of $p_0/\xi$.

\bigskip
\centerline{\it 3.2. Dispersion curves: $n_r(k)$}
\bigskip

The hydrodynamical system described by eqs. (2.1)-(2.5) admits four different
normal modes when the physical variables are perturbed as in equation
(2.6). The characteristic equation  is a fourth
degree polynomial with real coefficients (eq. 2.7) and has
zero or an even number of real roots. In other words, if there are
complex roots, they should be complex conjugate pairs.
In addition to the three modes found by Field (1965), a new mode
is introduced by the ionization equation (2.3). Investigation of
this additional ``reactive mode'' is important for two reasons: first,
because it may initiate an instability driven by ionization changes; second,
because the characteristics of the thermal mode might change.

The nature of the modes can be easily determined by studying their
behavior in the
small and large wavenumber limits and they can be classified as
two acoustic (hereafter denoted by WW)  and two thermo-reactive (TR) modes.
The solution in the limit of small (isochoric) and large (isobaric)
wavenumbers can be written as:

$$k\rightarrow \infty \left\{\eqalign
{n_1=&ic_sk-{b_1-b_3\over 2},\cr
n_2=&-ic_sk-{b_1-b_3\over 2},\cr
n_3=&{1\over 2} \Biggl\lbrack-b_3+\sqrt{b_3^2-4b_4}\Biggr\rbrack,\cr
n_4=&{1\over 2}\Biggl\lbrack-b_3-\sqrt{b_3^2-4b_4}
\Biggr\rbrack;\cr}\right.\eqno (3.1)$$

$$k \rightarrow 0 \left\{\eqalign
{ n_1=&ic_sk\sqrt{b_4\over
b_2}-{c_s^2k^2(b_2b_3-b_1b_4)\over 2b_2^2},\cr
n_2=&-ic_sk\sqrt{b_4\over b_2}-{c_s^2k^2(b_2b_3-b_1b_4)
\over 2b_2^2},\cr
 n_3=&{1\over 2}\Biggl\lbrack-b_1+\sqrt{b_1^2-4b_2}
\Biggr\rbrack,\cr
 n_4=&{1\over 2}\Biggl\lbrack-b_1-\sqrt{b_1^2-4b_2}
\Biggr\rbrack,\cr}\right.\eqno (3.2)$$

\noindent
where the coefficients $b_i$ are independent of $k$ and are related to
$a_i$'s of eq.(2.7) through a combination of powers of
$c_sk\equiv\tau_d^{-1}$

$$a_1=b_1;~~~ a_2=b_2+c_s^2k^2;$$
$$ a_3=b_3 c_s^2k^2;~~~ a_4=b_4c_s^2k^2.$$

The solutions $n_1$ and $n_2$
represent the acoustic modes: if not overdamped, they oscillate
with group velocity $dn_i/dk\sim c_s$. In the isochoric regime
propagation can be slower than the sound speed by a factor
$(b_4/b_2)^{1/2}$, which in all the
cases studied never differs from unity by more than a factor
of 10. These oscillatory solutions are always stable for $k=0$.

The thermo-reactive modes correspond to solutions
$n_3$ and $n_4$; $dn_i/dk=0$ for any $k$ and therefore
they do not propagate.
Since  in our case $b_1>0,b_2>0$, each equilibrium discussed
in this paper will be thermo-reactively stable at small $k$.
In this limit thermo-reactive modes are fast (in the sense that they are damped
in a time much smaller than the dynamical time)
and obey the reduced characteristic (isochoric) equation $n^2+b_1n+b_2=0$.
In the opposite limit (large $k$) thermo-reactive modes
grow or damp on timescales much longer than the dynamical time,
and satisfy the reduced (isobaric) equation $n^2+b_3n+b_4=0$.

In many cases a continuous dispersion curve $n_r(k)$
corresponds to an acoustic mode in the isochoric regime,
but to a thermo-reactive mode in the isobaric one.
The nature of each solution in this case is not conserved
when passing from the isochoric to the isobaric regime, and modes
might not be identifiable at some intermediate wavenumber.

We will illustrate in Figure 2 three different dispersion curves for a
particularly representative case, namely $E=100$~eV, $Z=1$, since
these summarize all the possible behaviors of $n_r(k)$ in
unstable regions. Numerical values refer to $N_0\rho=1$~cm$^{-3}$;
since wavenumber and growth rates scale linearly with the density,
extension to other density values is obtained multiplying both
$k$ and $n$ by $N_0\rho$.
All three curves represent WW modes for small $k$ (\ie $k \simlt
10^{-18}$~cm$^{-1}$) and TR modes for large $k$; TR (WW) modes
in the isochoric (isobaric) regime are stable for all three cases. In the
isochoric regime WW modes are oscillatory and unstable,
implying, according to eq. (3.2), $b_1b_4>b_2b_3$.
For $T_0=120$~K,  $n_r$ has a maximum
at $\log k=-17.8$ followed by an abrupt break leading to stable TR modes.
The minimum growth time  for WW  is $~ 2.5\times 10^7 (N_0\rho)^{-1}$~yr
(the curve shown has been multiplied by a factor of 10).
Increasing the temperature, TR modes become unstable for large $k$,
and the curve for $T_0=160$ K represents a transition case between TR
stable and unstable regimes. Except for this transition region,
when TR modes are unstable, their growth rate is usually
larger than the WW one (WW modes remain unstable anyway), and therefore
they represent the most dangerous modes for the system. A more typical
dispersion curve for unstable WW and TR modes is that shown for
$T_0=200$ K; as for the usual thermal instability $n_r$ grows
monotonically with $k$ and the curve flattens out as
$k\rightarrow\infty$. The effect of thermal conduction in this case
would be to stabilize very large wavenumber perturbations

\bigskip
\centerline{\it 3.3. Nature and properties of the modes}
\bigskip

It is possible to gain some qualitative understanding of the main
results from an analysis of the dispersion curves in the limit of
small and large $k$.
The physical explanation of the acoustic instability at small $k$
is the following.
When a fluid element is compressed and then rarefied by an acoustic wave
there will be a net energy flow into the wave, destabilizing it, if more energy
is gained by the gas during the compression phase than is lost in the expansion
phase (Field 1965). For long enough wavelengths, the dependence of the
ionization
on $\rho$ and $T$ brings this about as follows. For long enough wavelengths,
the
ionization time is shorter than the period of the wave, so we are in ionization
equilibrium at all times. During the compression phase $\rho$ increases,
increasing the recombination rate. $T$ may increase somewhat also, but not
enough to affect the recombination rate significantly. Hence, $x$ decreases in
the compression phase; since the cooling rate is proportional to $x$, it
decreases, and there is therefore a net gain of energy from the external
radiation field. In the expansion phase, $x$ increases, and there is a net heat
loss, but that is secondary, as shown by Field (1965), so the net effect of the
ionization degree of freedom is to destabilize long waves
(Flannery $\&$ Press 1979). If the wavelength is short enough so that the
ionization time is longer than the wave period, $x$ does not change, and the
wave mode becomes stable.
As a consequence, the existence of the critical wavenumber
for WW stabilization is fixed by the condition that the dynamical
time is roughly equal to the ionization time.
At that value of $k$ the oscillation rate ($\propto k$ in the
entire range) becomes
much shorter than the ionization time and stiffening
of the fluid caused by ionization is no longer possible.
Unstable acoustic waves are not found to be oscillatory
in the usual thermal instability study (see Fig. 1 of Field [1965]),
but they are overdamped if the dependence of the cooling function
on density arises from two-body collisions.

For $T_0=120$ K Figure 2 shows that the system is acoustically unstable
but thermo-reactively stable;
however, except for the effects of ionization,
it would be to thermally unstable at this temperature.
For the equilibrium conditions we have considered
{\it ionization is in fact a stabilizing agent} with respect to thermal
instability. In addition it may introduce an oscillatory part in the growth
rate,
and in this case we shall refer to them as overstable modes.
These findings have important implications and represent completely new
features introduced by ionization.

In the isobaric limit, we have seen at the beginning of this section that
the system (2.1)-(2.5) reduces to $ n^2+b_3n+b_4=0 $ for thermo-reactive
modes. This can be interpreted as the characteristic equation of a damped
oscillator described by
$$  {d^2y\over dt^2} +b_3 {dy\over dt}+ b_4 y=0,$$
where $y$ is the displacement of $x$ or $T$ from their equilibrium values,
$b_3$  is the damping coefficient and
$\sqrt b_4$ is the natural frequency of the oscillator.
A necessary condition both for stability (see also Hurwitz's criterion) and
for the motion to be oscillatory is $ b_4>0$.
Moreover, if $b_4>0$ stability depends only on the sign of $b_3$;
since from eq. (2.7) the terms $J_T-J_\rho$ and $-(J_x-J_\rho)$ in $b_3$ are
positive,
they both tend to make $ b_3>0$, hence to stabilize TR modes.

To understand the physical reason for the stabilization effect
suppose that the cooling is dominated by electron-metal impacts.
Since for large $k$ the perturbation is isobaric, a decrease in $T$
corresponds to an increase in the density, and, if the initial state is
thermally unstable, $\rho$ will increase even further.
However, on the ionization time scale (if not too large, see below)
$x$ will react to the increased $\rho$ by decreasing.
If $K_x-K_\rho>0$  the cooling will decrease, and as a consequence
the temperature will raise up again, trying to counteract the instability.
The condition  $K_x-K_\rho>0$ is always verified
when cooling is dominated by electron-metal impact and $x\ll1$,
since in that case the heat-loss function depends linearly on $x$, however
the end result of the perturbation depends also on time scales.
 From the expression of $b_3$ it can be seen for example that stabilization
by ionization is not effective for oscillatory modes if the characteristic
cooling
time scale $\tau_\rho\tau_T/(\tau_T-\tau_\rho)$ is  shorter than the
reactive timescales $\bar\tau_\rho\bar\tau_T/(\bar\tau_T-\bar\tau_\rho)$,
and $\bar\tau_\rho\bar\tau_x/(\bar\tau_\rho-\bar\tau_x)$.

\vglue 0.1 true in

\section
{NUMERICAL RESULTS: LINEAR PHASE}

\vglue 0.1 in

In this Section we present results of the linear stability
analysis. There are three free parameters in our model,
namely, the mean photon energy, $E$, the metallicity of
the gas, $Z$, and the photoionization rate $\xi$ (or, equivalently,
$T$). Over a wide range of wavenumbers and for each value of
$E,Z,\xi$, we solve numerically eq. (2.7) obtaining the four eigenvalues
and the instability domains.

If for given value of the parameters each eigenvalue has a positive real
part, $n_r<0$,for any $k$, the equilibrium is stable.
When this does not happen it is necessary to inspect the dispersion
curves $n_r-k$ to understand the specific nature of the instabilities
since isobaric and isochoric limits are known (eq.(3.1) and (3.2)).
In addition, dispersion curves provide information on the maximum
positive real part of the growth rates, $n_m$, which indicates the
most dangerous instability for the system.
Results are shown in Figure 3$(a)$ both for low-metallicity systems
and metal-rich ones. If more than one mode is unstable,
we plot only $n_m$ of the fastest one. If $n_r$
does not grow monotonically with $k$ towards some positive constant
but presents a maximum located at some wavelength $k_m$, in
Figure 3(b) we plot $k_m$.
Both wavenumber and growth rates refer to the case $N_0\rho=1$ cm$^{-3}$
and scale linearly with the density.

Figure 4 summarizes the extension of stable and unstable regions in the
plane $E-T$ for several values of $Z$.
Dotted areas indicate where only acoustic
instabilities exist and therefore these are the fastest modes.
For low metallicity $Z\simlt 0.1$ such regions do not exist; they
first appear at low $E$. For $Z\ge 1$ two separate branches which
are only acoustically unstable are found; one at low $T$ and one at high
$T$, each of which widens as $Z$ increases. For $E$ just above 13.6 eV
equilibria are more often thermally stable but acoustically unstable.
In the dashed regions the equilibrium
is unstable for any $k$: except close to the low and high temperature
borders next to the acoustic unstable regions,
the most unstable modes are the thermo-reactive
ones for large $k$. Therefore we shall refer to these regions as thermally
unstable regions. The extension of such regions is not a monotonic function
of $Z$: for the cases shown its minimum extension is for $Z=0.5$.
At these intermediate $Z$ the reactive mode has the strongest stabilizing
effect, while at high $Z$ it stabilizes the thermal mode but it produces
acoustically unstable modes.

\bigskip
\centerline{\it 4.1. Low-metallicity systems}
\bigskip

We examine here the stability of systems with a low metal content;
in particular we select the two values $Z=0$ and $Z=10^{-3}$.
In the absence of metals the stability analysis shows that
the hydrogen gas is {\it stable} for perturbations of all wavenumbers,
independently of the ionizing photon energy and of the intensity of the
ionizing photon flux.

A small amount of metals in the system  changes the shape of the
equilibrium curve, as already mentioned when discussing Fig. 1, and
an unstable mode appears for $100 \simlt T \simlt 6000$ K.
The mode is unstable at all wavenumbers and changes from  an
overdamped acoustic instability in the isochoric limit (small $k$) to a
thermo-reactive one for large $k$ where the largest growth rate occurs;
$n_i=0$,
for any $T,k,E$.

\bigskip
\centerline{\it 4.2. High-metallicity systems}
\bigskip

The nature of the instability and the unstable temperature
range in high metallicity systems depend more strongly on the
mean photon energy and the metal abundance. As
already mentioned in Section 3, there is one common feature to
unstable modes: either they are thermo-reactively stable at small scales,
$\lambda \simlt 1/(N_0\rho)$ pc, but acoustically overstable at larger
scales or they are unstable at all scales, acoustically at large scales but
thermo-reactively at small scales, where they mostly have the fastest
growth rates.

For low values of $E$ (see Figure 4) the gas is thermo-reactively stable
for $Z=0.5$ or lower but there are two overstable acoustic waves ($n_r>0$,
$n_i\ne 0$) over a wide range of $T$. For example, in the case of $E=15$~eV,
Figure 3 shows that the shortest growth time for these unstable acoustic modes
is for $T_0\simeq 500$~K and equals $2\times
10^6/(N_0\rho)$~yr for $Z=0.5$ and  $10^5 /(N_0\rho)$~yr
for $Z=1$. The scales corresponding to these growth times are
$\sim0.5/(N_0\rho)$~pc and $\sim0.1/(N_0\rho)$~pc respectively.
These parameters for $Z=1$ yield an optical depth $\sim 1.5$: this means
that the mean free path of photons is shorter than the wavelength of the
maximum   growth rate  ($2\pi/k_m$) and therefore it is likely that
the instability grows slowly on the mean free path scale, $N_0\rho\sigma(E)$,
if this is still unstable ($\sigma(E)$ is the hydrogen photoionization
cross section).
By increasing the mean photon energy the gas tends to become unstable
over all wavenumbers and the maximum growth rate is for TR modes with
$\lambda < (0.01/N_0\rho)$~pc.
For $E\gg 13.6$~eV, $Z$ needs to be larger than 0.5 to have
some range of temperatures where only acoustic modes are unstable.
For these energies  Fig. 4 shows that instabilities occurs independently
of $E$, and therefore at any depth inside the cloud.

The range of temperatures for which the gas is
thermo-reactively unstable becomes wider as we increase $Z$ from 0.5
to 1 or 2, independently of $E$.
The growth time of unstable modes decreases with the gas
metal content; for example Figure 3 shows that when $Z$ changes from 0.5
to 1 and $E\gg 13.6$ eV, the growth time becomes smaller by a factor
of 3. Notice that when TR modes are overstable the maximum growth rate
decreases: in Figure 3 for $Z=0.5$ and $E=40$ eV, $n_m$ shows a dip
in the interval $1000\simlt T\simlt 2000$~K, where there are
two overstable modes.
The frequency of oscillation $n_i$ is equal or higher than $n_r$.
For $Z\sim 1$ or larger and  temperatures close to the region where only
WW modes are unstable, TR modes are coupled and
oscillatory. The coupling between TR modes tends to vanish
as we increase $E$ or decrease $Z$.

\bigskip
\centerline{\it 4.3. Comparison with previous results }
\bigskip

In order to better understand the results discussed here it is instructive
to compare them with the ones obtained in previous works on the subject.

Field (1965) analyzed the thermal instability in a gas neglecting
fractional ionization changes. This implies that there are only three
modes, two waves and a non-oscillatory ``condensation'' mode.
If we apply Field's criterion to our model,
we find that there is an unstable non-oscillatory mode which
represents an acoustic overdamped wave at low $k$, while it becomes an
unstable condensation mode for large $k$ where the growth rate of the
mode is much larger. The main differences from Field's
analysis for a perturbation of the fractional ionization are:
$(a)$ the temperature range in which a gas is thermally stable is
wider and for low $E$ the thermal instability is suppressed for any $T$
(see Fig.~3-4); $(b)$ stability is no longer connected with the shape
of the $p_0/\xi$ curve; $(c)$ acoustic waves as well as thermal modes
are often oscillatory; $(d)$ there are regions where the gas is
thermally stable but nevertheless acoustic waves are
isochorically unstable.

Flannery \& Press (1979) already found
the presence of unstable acoustic waves in cold thermally stable regions.
They studied the stability of a photoionized gas
with a single metal coolant excited by electron impact only and under
the assumption $x\ll 1$. When the metal coolant is CII, $Z=1$, and
$E\simeq40$~eV, they find that unstable acoustic waves exist only
where the gas is thermally stable and $30<T<90$~K. With the more complete
cooling function adopted here (eq. 2.9) and dropping the condition
$x\ll1$, we find (see Figure 4 for the case $Z=1$) that:
$(a)$ the region $90<T<160$~K is no longer thermally unstable since ionization
is able to suppress the condensation mode for some equilibrium,
leaving acoustic waves as the only relevant unstable modes of the system;
$(b)$ the acoustically unstable region is located at $60<T<160$~K;
$(c)$ a new acoustically unstable region arises in the warm phase
for $T\simeq 8000$~K, a fact that may have several
interesting astrophysical consequences.

Defouw (1970) studied the stability of a pure hydrogen gas cooled
by recombinations only and including the reactive mode.
He mainly focused on a collisionally ionized
gas, dealing with the photoionized case only briefly. Apart from an
incorrect recombination coefficient (see \S 3), he stated that
photoionized gas is thermally unstable if the mean energy of the photoelectron
is less than the mean kinetic energy of the recombining electron. In our
notation, this condition translates into $E-{\rm \chi}< kT$, and therefore
cases with $E\gg 13.6$~eV should be stable. However, if heating is
provided by photons with $E\gg 13.6$~eV the system cannot be in equilibrium
if the only cooling is provided by radiative recombinations; thus,
according to Defouw, the gas should be always unstable (Yoneyama 1973
has studied a model similar to the one by Defouw, reaching similar
conclusions). Inclusion of e-H collisional excitation cooling allows
several stable equilibrium
temperatures. A detailed model for a collisionally ionized pure hydrogen
plasma, with the inclusion of e-H cooling, has been studied by Iba\~nez
\& Parravano (1983) who obtain a modified instability criterion with
respect to Defouw (the gas is stable in the two temperature ranges
$T< 8000$~K and $13000 {\rm ~K} < T < 19000 {\rm ~K}$ while for Defouw the
gas is stable for $T<17000$ K). However these authors have not studied the
stability for the case of a photoionized, metal cooled gas and therefore
their results are not directly comparable with ours.

\vglue 0.1 true in

\section
{NONLINEAR ANALYSIS}

\vglue 0.1 in

The aim of this Section is to investigate the nonlinear development of the
TR instability in the large $k$ limit by solving the
complete set of hydrodynamic equations for the isobaric case numerically.
We choose an equilibrium state $\zeta_0\equiv(T_0, x_0)$ (which also fixes
the value of $p/\xi$ uniquely) and impose a nonlinear perturbation
$\delta\equiv (\delta T, \delta x)$; the resulting perturbed state
$\zeta=\zeta_0+\delta$ is then used as the initial condition of the problem.
Typical amplitudes of the perturbations are below 5\% of the equilibrium
value, and we performed several runs to test that the final result is
insensitive
to the values and signs of $\delta T$ and  $\delta x$. We did not take higher
values of $\delta$
since we believe that stronger initial fluctuations are of dubious physical
origin. One may wonder if the isobaric assumption for the evolution is perhaps
too severe and motions could develop; this requires a full
hydrodynamical calculation. In this context we limit ourselves to solve
the isobaric limit of the energy and ionization equations using a standard
Runge-Kutta solver.

The most interesting questions concern the fate of unstable equilibria
for which there are no corresponding stable temperatures for the
same $p_0/\xi$. Important points are also the timescales of the formation
of condensations usually originated by thermal instabilities.
Before going into the detailed description of a representative case, we
briefly summarize the main results of the nonlinear analysis.

1. The final fate of the nonlinear perturbation depends on the availability
of {\it stable} states at the same pressure. If there are such states,
the system will make a phase transition, as one could have expected.

2. When unstable states do not have other isobaric stable states (a
situation not possible if ionization is considered constant) the system
enters an oscillating, periodic state without approaching any stable value.

3. Oscillations, if present in the linear or nonlinear phase, cause a much
slower transition to the final state.

For sake of clarity we will discuss separately unstable {\it multi-phase}
(\ie two or more isobaric equilibria of which
at least one is stable) or unstable {\it single-phase} (\ie one or more
isobaric
unstable equilibria). When the photon energy is high ($E\gg 13.6$~eV)
unstable equilibria are multi-phase; in the opposite case they are always
single-phase (see Fig. 1).
To illustrate in detail the results summarized above, we take an
intermediate case: $E=40$~eV, $Z=1$, which  has both behaviors depending on
the value of $p_0/\xi$. This being a transition case between monotonically
growing curves of $p_0/\xi$ at low $E$ and wide multi-phase curves at
high $E$, the shape of the curve is a bit peculiar and the
range in which the $p_0/\xi$ curve allows for multi-phase equilibria
is rather narrow. However, as discussed by Krolik \etal (1981) small ranges
of $p_0/\xi$ involve wide range of gas densities and it is not so unlikely
for the gas to lie in these regions. The equilibrium curve has been expanded
for convenience in Fig. $5(a)$ and four points are marked in it for which
nonlinear results are shown in Fig. 5$(b)$ and discussed below.
Numerical results refer to perturbations of equilibria with
$N_0\rho=1$~cm$^{-3}$.

The first point $P1$ is a thermo-reactive stable equilibrium with
$T_0=120 {\rm ~K}, x_0=0.012$. In the linear phase the perturbation shows
oscillations with $n_i=5\times 10^{-6}$~yr$^{-1}$. These oscillations are
clearly seen also during the nonlinear evolution before the perturbation
is damped, a process which takes about $400 \tau_h=10^7$~yr.

$\bullet$ {\it Single-phase equilibria}. Equilibria $P2$ and $P3$ in Fig.
5$(a)$ ($E=40$~eV) correspond to single-phase equilibria.
For $P2$, $T_0=150$~K and $x_0=0.014$,  and the perturbation is overstable in
the linear
phase (see also discussion in Sec. 3.2) with oscillation rate larger than
the growth rate. For this reason, a large number of oscillations of growing
amplitude is seen in $T,x,\rho$ before saturation takes place (Fig. 5$(b)$),
as usually happens for thermo-reactive unstable
equilibria close to the unstable/stable transition region. After
$t\sim 10^3\tau_h$ the system relaxes to a state of {\it periodic}
oscillations of constant amplitude. Note that the ionization lags the
temperature, as expected.
Far from the unstable/stable transition region, as for the equilibrium point
$P3$ ($T_0=1000$~K, $x_0=0.075$) the value of $n_i$ is, in
contrast, zero or rather small compared to the to the growth rate.
Therefore for $P3$ the perturbations grow to the saturated values in less
than one oscillation period. After that moment, the evolution is
qualitatively similar to that of $P2$ but the oscillation is not centered
around $T_0$, and the
temperature spans the very large range $1000 {\rm ~K}\simlt T \simlt 10,000$~K.

$\bullet$ {\it Multi-phase equilibria}. The point labelled $P4$ in Fig. 5
corresponds to an equilibrium temperature $T_0=2700 {\rm ~K}$,and fractional
ionization $x_0=0.18$. This equilibrium, which  in the linear phase is
thermo-reactively unstable with $n_i=0$, is obtained
when $\xi\simeq 3\times 10^{-14}$~s$^{-1}$ for $N_0\rho=1$~cm$^{-3}$.
The same value of $p_0/\xi$ corresponds also to the stable equilibrium point
$T_0=8100 {\rm ~K}, x_0=0.44$ but with a value of the volume density about 3
times
smaller. The full nonlinear evolution of temperature and ionization
fraction at constant $p_0/\xi$ is plotted in Fig. 5$(b)$ as a
function of time  (normalized to the heating time $\tau_h$).
The system reaches the other stable equilibrium after about
600$\tau_h=7.5\times 10^6$~yr through (nonlinear)
oscillations. This time is much longer than the typical time scale
that one would have calculated without the ionization, and, indeed, the
system spends a long time in a time-dependent, nonequilibrium state.
This retarding effect, due to ionization and already
discussed for the linear stage, occurs very often and an additional,
slightly different, case is shown in Fig. 6, corresponding to $E=100$~eV,
$Z=1$. Both temperatures, $T=150$~K and $T=1000$~K, correspond to
multi-phase equilibria; but when perturbed the first one shows oscillations
in the linear phase  whereas for the second $n_i=0$. It is clearly seen
in the enlarged parts (bottom panels of Fig. 6) that the characteristic
time for the phase transition is longer when oscillations are present.

To conclude, isobaric TR instabilities produce much slower phase
transitions than the usual thermal instabilities. When no other
stable equilibrium is available the medium spends long time
intervals in  nonequilibrium states characterized by nonlinear
periodic oscillations of temperature, density and ionization fraction.

\vglue 0.1 true in

\section
{SUMMARY AND IMPLICATIONS }

\vglue 0.1 in

We have studied the acoustic and thermo-reactive instabilities
in a diffuse
ISM heated and ionized by an external radiation field, of mean photon
energy $E$ and photoionization rate $\xi$, cooled
by collisional excitation of metals and hydrogen lines.
Thermo-reactive modes are found when the effects of time dependent
ionization on the standard thermal instability are taken into account.
The addition of
the hydrogen recombination reaction  to Field's classical
treatment of thermal instability has several remarkable consequences
especially when
the ratio between the heat input and the ionization rate of the radiation
field is rather small. In addition to the linear stability analysis we
have investigated the nonlinear development of thermo-reactive modes.
In general
it would be interesting to extend this type of  analysis to additional
ionization and heating mechanisms (decaying neutrinos, dust grains, Alfven
waves dissipation). For example, when present, dust grains may
be important heating sources via the photoelectric effect.
Since in this case the
heat input depends linearly on the gas density, as for the photoionization
heating, and the energy input per photoelectron is quite low ($\sim 1$~eV),
it is likely that some of the effects we found for low mean photon energies
apply, if hydrogen remains the main free electron contributor.
We summarize below the main results obtained in the
range of temperatures of interest, $30{\rm ~K}<T<30,000$~K and for metallicity
in the range $0\le Z \le 2$.

The stability properties depend weakly on the
mean photon energy absorbed:
from Fig. 4 it is evident that as we penetrate inside the gas
cloud, thus increasing $E$, the unstable modes remain the same.
This means that the stability of the gas is mainly governed by the temperature
rather than by the details of the radiation spectrum.
Only for photon energies close to the Lyman limit and for intermediate values
of
$Z$ the response of the medium to perturbations may depend on $E$,
and therefore edges of irradiated clouds can be sensibly different from
their interiors.

In the absence of metals the gas is stable.
A very small amount of metals induces non oscillatory unstable
acoustic and thermo-reactive modes and, as in
Field's analysis, the fastest growing instability is the isobaric
thermal mode. Unstable equilibria require $p_0/\xi\simgt
10^{17}/Z$~cm$^{-3}$~K~s, and growth times are short compared to the
Hubble time only if $N_0\rho Z\simgt 10^{-3}$~cm$^{-3}$.
Low metallicity models can apply, for example, to low redshift
Ly$\alpha$ clouds. In this case $p_0/\xi$ values are sufficiently
low that only a stable warm phase could persist. According to Defouw's
photoionization model  (where no e-H cooling has been considered)
these optically thin objects should instead be thermo-reactively unstable.

If the metal abundance is about half solar or higher the linear analysis
shows that unstable modes often become oscillatory and ionization is
a stabilizing agent, as explained in the  previous Sections.
Even though for column densities above $10^{18}$ cm$^{-2}$,
where the mean photon energy of the penetrating flux is larger than the
hydrogen ionization energy, the range of thermally unstable temperatures is
still wide, at low energies this range is considerably smaller and growth
times are longer. Furthermore, a new type of situation arises in which the
gas is thermo-reactively stable but unstable acoustic waves develop on
large scales. While for $Z\simeq 0.5$ this occurs only for
optically thin media, with $E-\chi \ll 100$ eV, for larger $Z$ two
regions are found which are only acoustically unstable, one in the cold
phase at $T\sim 100$~K and one in the warm phase at $T\sim 8000$~K.
In places where there
is no star formation for example, as in outer galactic HI disks,
the development of these unstable acoustic waves could be the main source
of  motions in the gas. A nonlinear analysis of the evolution of the
unstable acoustic waves is however necessary to assess whether shock waves
develop and produce observational effects like non-thermal line broadening.

Another new and important feature introduced by the hydrogen
recombination reaction
is that the stability of isobaric thermo-reactive modes is no longer
connected with the shape of the $p_0/\xi$ curve. This implies that there
are unstable equilibria which cannot make an isobaric phase transition to
any other stable state. The nonlinear analysis has shown
that in this case the fate of the gas is to evolve towards a nonequilibrium
state characterized by periodic, nonlinear oscillations of density,
temperature and hydrogen ionization fraction.
This effect is especially important for small heat input/ionization rate
ratios, and for $Z>0.1$.

Promising objects to look for nonequilibrium oscillatory states are
clouds located outside the optical disk of galaxies, or in
the halo, like HVCs. These objects usually have low column
density and therefore relatively low energy photons can penetrate.
Their metallicity, though lower than solar, is still appreciably high
(de Boer \& Savage 1984).
Let us examine for example clouds with HI column density
$N_H\simeq 10^{18}$~cm$^{-2}$ irradiated by an extragalactic
background flux with spectral index 1.5 as
suggested by Sargent \etal (1979) and Madau (1992). The mean energy
of ionizing photons at half total column density above the center
is $E\sim 20$~eV. Taking the value $p_0\simeq100$~cm$^{-3}$~K as
speculated by Ferrara \& Field (1994), with a flux
intensity $I_\nu \sim 6\times 10^{-24}$~erg~cm$^{-2}$~s$^{-1}$~
Hz$^{-1}$ at the Lyman continuum, the corresponding ionizing rate
is $\xi\simeq 2\times10^{-15}$~s$^{-1}$ and $p_0/\xi\simeq
5\times 10^{16}$~cm$^{-3}$~K~s. With these parameters the
equilibrium temperature is $T_0\simeq 800$ K and a single-phase
thermo-reactive mode develops. The temperature in the
nonlinear stages oscillates
between 200 and 2500 K with a period of about $10^8$ yr, after
a slightly longer nonlinear growth time. The average temperature
corresponds to a FWHM of the HI line $\simeq 8$~km~s$^{-1}$.

There are other environments where ionizing photon energies can be low
enough that the nonequilibrium picture derived here can be expected,
in contrast with the predictions of two-phase ISM models.
Examples are high galactic latitude neutral hydrogen, where in fact
average temperatures derived from observations correspond  to thermally
unstable equilibria (Verschuur \& Magnani 1994), and interfaces between low
density HII regions and neutral gas.
Even if the results derived here may not be
directly applicable to bright HII regions where the important radiating
ions are not those considered in this paper, photoionization models
predict that low ionization
stages of metals dominate the cooling in the diffuse ionized layer
responsible for the pervasive Galactic H$\alpha$ emission
(see for example Domg\"orgen $\&$ Mathis 1994). When photons
from the stars penetrate through interstellar ``tunnels'', the gas can be
in a nonequilibrium state involving isobaric periodic oscillations, or
show motions due to unstable acoustic waves, depending upon local values of
optical depth and thermal pressure.

\vfill\eject

\baselineskip=10pt
\vsize=8.52 true in
\hsize=5.89 true in
\voffset=-0.037 true in
\hoffset=0.2 true in

\vglue 0.3 true cm
\hglue 8.5 true cm
{TABLE 1}

\vglue 0.1 true cm
\noindent
\hglue 7 true cm
SYMBOLS USED FREQUENTLY
\settabs\+012345678901234567890&0123456789012345678901234567890\cr 
\vglue 0.2 true cm
\hrule width20 cm
\hrule width20cm
\vskip 0.04 true cm
\hrule width20cm
\hrule width20cm
\vskip 0.2 true cm
\+Symbol & Meaning \cr
\+ \ (1)&(2) \cr
\vskip 0.2 true cm
\hrule width20cm
\vskip 0.2 true cm

\+$\rho,T,{\bf v},p,x$   & Density, temperature, velocity, pressure and
ionization fraction;\cr

\+ & subscript 0 indicates unperturbed value, 1 indicates
perturbation\cr

\+ $\gamma, N_0, k_B$ & Ratio of specific heats, Avogadro number,
Boltzmann constant \cr

\+ Z & Metallicity \cr

\+ ${\cal L}, {\cal L}_T, {\cal L}_\rho, {\cal L}_x$ & Generalized heat-loss
function per unit mass
and its T- $\rho$- and $x$- derivatives \cr

\+ $I, I_T, I_\rho, I_x$ & Generalized ionization-recombination function
per unit mass and its T- $\rho$- and $x$- derivatives \cr

\+ $\Lambda_{HZ}, \Lambda_{eZ}, \Lambda_{eH}, \Lambda_{eH^+}$ &
Cooling losses due hydrogen-metal, electron-metal, electron-hydrogen,
electron-proton impact\cr

\+ $c_s$ & Speed of sound \cr

\+ $n, k, \lambda$ & Growth rate, wavenumber, and wavelength of perturbation
\cr

\+ $n_r, n_i$ & Real and imaginary part of the growth rate $n$ \cr

\+ $n_m, k_m $ & Largest real part of the growth rates, and
value of $k$ for which this maximum occurs \cr

\+ $\xi, E$ & Photoionization rate and mean photon energy of the
incident radiation at a given optical depth\cr

\+ $\phi, E_h$ & Number of secondary electrons and thermal heat released
per photoionization\cr

\+ $\tau_h$,$\tau_d$ & Heating time, $\tau_h\equiv (5/2) k_BT/\xi E_h$,
dynamical time, $\tau_d\equiv 1/c_sk$ \cr

\+ $\tau_T$,$\tau_\rho$,$\tau_x$ & Characteristic cooling times for the
perturbed gas, where: \cr

\+ & $\tau_T \equiv (3/2)k_B T (1+x) N_0/|T {\cal L}_T|$,
$\tau_\rho \equiv (3/2)k_B T (1+x) N_0/|\rho {\cal L}_\rho|$, $\tau_x \equiv
(3/2)k_B T (1+x) N_0/|(1+x){\cal L}_x|$ \cr

\+ $\bar\tau_T$,$\bar\tau_\rho$,$\bar\tau_x$, & Characteristic reactive times
for the
perturbed gas, where: \cr

\+ & $\bar\tau_T \equiv (1+x) N_0/|T I_T|$, $\bar\tau_\rho \equiv
(1+x) N_0/|\rho I_\rho|$,
$\bar\tau_x \equiv (1+x) N_0/|(1+x) I_x|$ \cr

\+ $K_T, K_\rho, K_x$ & Characteristic wavenumbers associated with
cooling processes:\cr

\+ & $K_T=({\cal L}_T/|{\cal L}_T|)/\tau_T c_s$, $K_\rho=({\cal L}_\rho/|{\cal
L}_\rho|)/\tau_\rho c_s$,
$K_x=({\cal L}_x/|{\cal L}_x|)/\tau_x c_s$ \cr

\+ $J_T, J_\rho, J_x$ & Characteristic wavenumbers associated with
reactive processes:\cr

\+ & $J_T=(I_T/|I_T|)/\tau_T c_s$, $J_\rho=(I_\rho/|I_\rho|)/\tau_\rho c_s$,
$J_x=(I_x/|I_x|)/\tau_x c_s$ \cr

\+ $\Theta$ & Inverse temperature $\equiv 157900/T$ \cr

\+ $\gamma_c$ & Collisional ionization rate\cr

\+ $\chi$, $\tilde\chi$  & Hydrogen ionization potential,
$\tilde\chi \equiv\chi/(3/2)k_B T$\cr

\+ $ \alpha, E^{th}$  & Total radiative recombination coefficient and thermal
energy lost by recombinations to all levels $ {\rm n}\ge 2$ \cr

\vskip 0.2 true cm
\hrule width20cm
\hrule width20cm
\vskip 0.2 true cm
\vfill\eject

\heading {ACKNOWLEDGMENTS}

We are grateful to J. M. Dickey, E.E. Salpeter and especially to G. B.
Field for interesting discussions and comments on the subject of this paper.
E.C. acknowledges the Agenzia Spaziale Italiana for financial
support.

\bigskip

\def\refindent{\advance\leftskip by 24pt \parindent=-24pt}

\heading {REFERENCES}

\vglue 0.1 in

\refindent
Aleksandrov, A.D., Kolmogorov, A. N. \& Lavrent'ev, M. A. 1963,
     Mathematics: Its Content, Methods, and Meaning, (MIT Press:~Cambridge)\par

Black, J. H. 1981, MNRAS, 197, 555\par

Charlton, J. C, Salpeter, E. E. \& Hogan C. J. 1993, ApJ, 402, 493\par

Collin-Souffrin, S. 1990, in New Windows to the Universe, vol. 2,
ed. F. Sanchez \& M. Vasquez (Cambridge:~Univ~Press), 235\par

Corbelli, E. \& Salpeter, E. E.  1993a, ApJ, 419, 94\par

Corbelli, E. \& Salpeter, E. E.  1993b, ApJ, 419, 104\par

Dalgarno, A. \& McCray, R. A. 1972, ARA\&A, 10, 375\par

de Boer, K. S. \& Savage, B. D. 1984, ApJ, 136, L7\par

Defouw, R. J. 1970, ApJ, 161, 55\par

Dickey, J. M., Murray, H. M. \& Helou, G. 1990, ApJ, 352, 522\par

Domg\"orgen, H. \& Mathis J. S. 1994, ApJ, 428, 647\par

Dove J. B. \& Shull, J. M. 1994, ApJ, 423, 196\par

Ferrara, A. \& Field, G. B. 1994, ApJ 423, 665\par

Field, G. B. 1965, ApJ, 142, 531\par

Field, G. B., Goldsmith, D. W. \& Habing, H. J. 1969, ApJ, 155, L149\par

Flannery, B. P. \& Press, W. H. 1979, ApJ, 231, 688\par

Goldsmith, G. W. 1970, ApJ, 161, 41\par

Iba\~nez, M. H. \& Parravano A. 1983, ApJ, 275, 181\par

Ikeuchi S. \& Turner 1991, ApJ, 381, L1\par

Krolik, J. H.,  McKee, C. F., \& Tarter, C. B. 1981, ApJ, 249, 422\par

Kulkarni, V. S. \& Fall, S. M 1993, ApJ, 413, L63\par

Kwan, J. Y. \& Krolik, J. H. 1981, ApJ, 250, 468\par

Launay, M. \& Roueff, E. 1977, A\&A, 56, 289\par

Madau, P. 1992, ApJ, 389, L1\par

Maloney, P. 1993, ApJ, 414, 14\par

McKee, C. F. \& Ostriker, J. P. 1977, ApJ, 218, 148\par

Miralda-Escud\'e, J. \& Ostriker, J. P. 1992, ApJ, 392, 15\par

Reynolds, R. J. 1993, in 3rd Annual Maryland Meeting, Back to the
Galaxy, (Washington:~NASA), 156\par

Sargent, W. L. W., Young, P. J., Boksenberg, A., Carswell, R. F.,
\&  Whelan, J. A. J. 1979, ApJ, 230, 49\par

Shull, J. M., \& Van Steenberg, M. E. 1985, ApJ, 298, 268\par

Songaila, A., Bryant, W. \& Cowie, L. L. 1989, ApJ, 345, L71\par

Yoneyama, T. 1973, PASJ, 25, 349\par

Verschuur, G. L. \& Magnani, L. 1994, AJ, 107, 287\par
\vfill\eject

\heading
{FIGURE CAPTIONS}

\vglue 0.1 in

{\bf Figure 1}. Equilibrium curves of $p_0/\xi$ for four values of the
metallicity $Z$. $p_0$ is in units of cm$^{-3}$ K and $\xi$ is the
photoionization rate in s$^{-1}$. For each value of $Z$ we show the
equilibrium curves for three energies: in each panel the bottom
curve is for $E=15$ eV, the middle curve is for $E=40$ eV and the
top curve is for $E=100$ eV. Dashed lines denote unstable equilibria;
solid lines denote stable equilibria.

{\bf Figure 2}. Positive real part of growth rates $n_r$ as
function of the wavenumber $k$ for three equilibrium temperatures relative
to $E=100$~eV, $Z=1$, and $N_0\rho=1$~cm$^{-3}$.
For $T_0=120$ K we plot the growth rate multiplied by a factor of 10.
For this temperature the equilibrium is stable at large $k$ while for
$T_0=160$~K and $T_0=200$~K there is an unstable mode at all wavenumbers.
The maximum growth rate for $T_0=200$~K is for $k\rightarrow\infty$.

{\bf Figure 3}. In $(a)$ the largest positive real part of growth
rates, $n_m$, is plotted as a function of the equilibrium temperature $T_0$.
In $(b)$ we plot the wavenumber $k_m$ for which the largest growth rate
has been found. For some values of $n_m$,  $k_m\rightarrow\infty$; in this
case $k_m$ has not been plotted. We have used $N_0\rho=1$ cm$^{-3}$.
Each panel in $(a)$ or $(b)$ refers
to a different photon energy $E$ (eV). The continuous curves are for
metallicity $Z=1$, the small dashed curves for $Z=0.5$ and the
dotted-dashed curve for $Z=10^{-3}$.

{\bf Figure 4}. Unstable regions in the plane $E-T$ for various
values of the metallicity $Z$. In the dotted regions there are
unstable modes only for $k<k_m$ (acoustic waves). In the dashed
regions equilibria have unstable modes for any $k$
(they are both acoustically and thermo-reactively unstable).
The dashed curves at the bottom and left hand side of
each panel limit the possible equilibrium values of $T_0$ for each
value of the photon energy $E$.

{\bf Figure 5}. $(a)$ part of equilibrium curve $p_0/\xi-T_0$
for $E=40$ eV and $Z=1$. The dashed-dotted lines refers to
equilibria which have unstable thermo-reactive modes.
For the four equilibrium temperatures marked on the curve
we plot in $(b)$ the nonlinear evolution of the
temperature and ionization fraction perturbations in the isobaric
regime.
Continuous curves refer to $\delta T/T_0$, dotted curves to
$\delta x/x_0$ as function of time (normalized to the heating
time $\tau_H$). The density is set to $N_0\rho=1$ cm$^{-3}$.

{\bf Figure 6}. Nonlinear evolution of temperature and ionization
fraction perturbations in the isobaric regime for two multi-phase
equilibria in the case $E=100$ eV, $Z=1$, and $N_0\rho=1$ cm$^{-3}$.
Continuous curves refer to $\delta T/T_0$, dotted curve to
$\delta x/x_0$ as function of time (normalized to the heating
time $\tau_H$). Bottom panels show enlarged portions of top
panels.

\vfill\eject
\end